\documentclass[12pt]{iopart}
\usepackage{times}
\usepackage{graphicx}

\begin{document}

\title{Dirac monopoles from the Matsumoto non-commutative spheres}
\author{Tomasz Brzezi\'nski\footnote{TB thanks the Engineering and Physical Sciences Research 
Council for an Advanced Fellowship.}}
\address{Department of Mathematics, University of Wales Swansea, Swansea SA2 8PP, UK}
\author{Andrzej Sitarz}
\address{Institute of Physics, Jagiellonian University, Reymonta 4, 30059 Krak\'ow, Poland}
\begin{abstract}
It is shown that the non-commutative three-sphere introduced by Matsumoto is a total space of the quantum Hopf  bundle over the classical two-sphere. A canonical connection is constructed, and is shown to coincide with the standard Dirac magnetic monopole.
\end{abstract}

\section{Introduction}
One of the first examples of a non-commutative three-sphere was constructed by Matsumoto in \cite{Mat:sph}. This example was overlooked for a considerable time, mainly due to a focused attention attracted by a plethora of examples of quantum spheres coming from quantum groups. The interest in Matsumoto's sphere was revived recently by
general revival of interest in non-commutative spheres triggered 
by papers \cite{ConLan:man}, \cite{DabLan:ins} and \cite{CoDuVi}, 
where  
the Matsumoto spheres were re-derived through the construction of projective 
modules with vanishing lower Chern classes and were shown to 
satisfy the axioms of a non-commutative manifold proposed by 
A.\ Connes. 

In a follow-up \cite{Mat:sph2} to his original paper, Matsumoto has also constructed a non-commutative Hopf fibering, thus making the first step toward developing gauge theory with non-commutative spheres. The aim of this note is to review Matsumoto's construction of non-commutative spheres, and then reformulate the Hopf fibering in terms of more modern language of quantum principal bundles as initiated in \cite{BrzMaj:gau}, in order to develop gauge theory on such spheres. In particular we prove that Matsumoto's Hopf fibering is a quantum principal bundle or a Hopf-Galois extension. We also construct a connection in this quantum principal bundle, and show that it coincides with the standard Dirac magnetic monopole. In other words we show that a classical physical particle can be equivalently described as a particle existing in a non-commutative world.
\section{Review of Matsumoto's construction}
The main idea of Matsumoto's construction is based on the deformation of the topological construction of the classical three-sphere via the {\em Heegaard splitting}. Classically, one starts with the solid two-torus $D^2\times S^1$. The boundary of $D^2\times S^1$ is the two-torus $S^1\times S^1$. A loop on $S^1\times S^1$ is called a {\em meridian} if it vanishes in the fundamental group of $D^2\times S^1$, and it is called a {\em longitude} if it generates this group. The three-sphere $S^3$ can be obtained by gluing two copies of solid two-tori through a homeomorphism which exchanges a meridian with a longitude on their common boundary. Algebraically, in terms of algebras of continuous functions this construction can be summarised as
$$
C(S^3) = \{(a,b)\in C(D^2\times S^1) \oplus C(S^1\times D^2) \; | \; \rho(\pi(a)) = \pi(b)\},
$$
where $\pi: C(D^2\times S^1)\to C(S^1\times S^1)$ is the canonical projection (restriction of the domain to the boundary), and $\rho: C(S^1\times S^1)\to C(S^1\times S^1)$ is an isomorphism corresponding to exchanging a meridian for a longitude.

Now, one can replace the algebra of functions on a solid torus, by a (non-commutative) $C^*$-algebra $D_\theta$ of functions on a non-commutative solid torus. This is generated by a normal operator $y$ and a unitary $v$ such that $vy = e^{2\pi i\theta}yv$, where $\theta$ is a non-zero real number. There is a $C^*$-algebra surjection $\pi_\theta$ from $D_\theta$ to the non-commutative torus $A_\theta$. The latter is generated by unitaries $U,V$ subject to the relations $VU = e^{2\pi i\theta}UV$. The surjection $\pi_\theta$ is given by $\pi_\theta(v) = V$, $\pi_\theta(y) = U$, and is the non-commutative version of the canonical projection $\pi$ above. Next, take $D_{-\theta}$ generated by a normal operator $x$ and a unitary $u$, and the corresponding non-commutative torus $A_{-\theta}$ generated by $\hat{U},\hat{V}$.  A $C^*$-algebra map 
$$
\rho_\theta : A_{-\theta}\to A_\theta, \qquad (\hat{U},\hat{V})\mapsto (V,U),
$$
is an isomorphism, which corresponds to the map interchanging meridians with longitudes. The deformed algebra of functions on the three-sphere can be thus defined  as
$$
C_\theta(S^3) = \{(a,b)\in D_{-\theta} \oplus D_\theta \; | \; \rho_\theta(\pi_\theta(a)) = \pi(b)\}.
$$
The algebra $ C_\theta(S^3)$ can be seen as generated by two normal operators $S=(u,y)$, $T=(x,v)$ and relations $TS = \lambda ST$, $(1-TT^*)(1-S^*S) = 0$ and $||S|| = ||T||=1$, where $\lambda = e^{2\pi i\theta}$. In \cite{MatTom:len}, Matsumoto and Tomiyama observed that, equivalently,  $C_\theta(S^3)$ is a $C^*$-algebra which has presentation with  generators $a,b$ and relations
\begin{equation}
aa^* = a^*a, \quad bb^* = b^* b, \quad ab = \lambda ba, \quad ab^* = \bar{\lambda} b^* a, \quad aa^* +bb^*=1.
\label{pres}
\end{equation}
The presentation (\ref{pres}) makes it clear that $ C_\theta(S^3)$ is a deformation of functions on the classical sphere, and it also allows one to identify the Matsumoto sphere with one of the spheres in \cite{ConLan:man}.

\section{Non-commutative Hopf bundle}
There is an action of $U(1)$ on the deformed $S^3$, which algebraically is represented as a {\em coaction} of a Hopf algebra (quantum group) $C(U(1))$ of functions on  $U(1)$ on $C_\theta(S^3)$. Explicitly, let $C(U(1))$ be generated by a unitary $Z$ (i.e., $ZZ^*=Z^*Z = 1$). Then the coaction $\Delta_R : C_\theta(S^3)\to C_\theta(S^3)\otimes C(U(1))$ is defined as a $*$-algebra map given on generators by
$$
\Delta_R(a) = a\otimes Z, \qquad \Delta_R(b) = b\otimes Z.
$$
The quantum quotient space under this (co)action is defined as a {\em subalgebra of coinvariants},
$$
C_\theta(S^3/U(1)) = C_\theta(S^2) := \{ f\in C_\theta(S^3) \; | \; \Delta_R(f) = f\otimes 1\}.
$$
One easily computes that $C_\theta(S^2)$ is generated by $z = aa^*$, $x_+=ba^*$, $x_- = ab^* $. This is a commutative algebra with an additional relation $z^2 + x_+x_- = z$, and thus coincides with the usual algebra of functions on the two-sphere, i.e., $C_\theta(S^2) = C(S^2)$. 

Note that the coaction of $C(U(1))$ on $C_\theta(S^3))$ defines a $\bf Z$-grading on the latter. An element $f\in C_\theta(S^3)$  has degree $n=0,1,2,\ldots$ if $\Delta_R(f) = f\otimes Z^n$, and has degree $-n$ if $\Delta_R(f) = f\otimes Z^{*n}$. Clearly $C(S^2)$ is a subalgebra of degree zero elements.

To make sure that the above construction defines a quantum principal bundle, one needs to check that the action represented by $\Delta_R$ is free. Algebraically, this means that we need to prove that the extension of algebras $C(S^2)\to C_\theta(S^3)$ is a {\em Hopf-Galois extension}, i.e., that the map ${\rm can} : C_\theta(S^3)\otimes_{C(S^2)}C_\theta(S^3) \to C_\theta(S^3) \otimes C(U(1))$ given by $f\otimes g\mapsto f\Delta_R(g)$ is bijective. Consider a map $\chi: C_\theta(S^3) \otimes C(U(1))\to C_\theta(S^3)\otimes_{C(S^2)}C_\theta(S^3)$,
$$
\chi (f\otimes Z^n) = \sum_{m=0}^n \lambda^{-m(n-m)}\pmatrix{n\cr\\ m}f a^{*n-m}b^{*m} \otimes a^{n-m} b^m,
$$
$$
\chi (f\otimes Z^{*n}) = \sum_{m=0}^n \lambda^{-m(n-m)}\pmatrix{n\cr\\ m}f a^{n-m}b^{m} \otimes a^{*n-m} b^{*m},
$$
for all $f\in C_\theta(S^3)$ and $n=0,1,2,\ldots$ One easily checks that ${\rm can}\circ \chi = \rm id$. The proof that also $\chi\circ {\rm can} = \rm id$ is also based on direct calculation. To facilitate this calculation one needs to make a couple of observations. First, since both maps are obviously left $C_\theta(S^3)$-linear, suffices it  to check the identity on elements of the form $1\otimes f$. Second,  note that the tensor product is over $C(S^2)$, hence one may restrict oneself to elements of type $f=a^mb^n$ with $m,n\geq 0$. Now the calculation is straightforward. Thus we conclude that $\chi$ is the inverse of $\rm can$, i.e., we have defined a quantum $U(1)$-principal bundle over $S^2$ with the total space given by the Matsumoto's three-sphere. This quantum bundle is denoted by $(C(S^2)\subset C_\theta(S^3))^{C(U(1))}$.

\section{Non-commutative setup for the Dirac monopole}
Since the Matsumoto three-sphere is a total space of a quantum (non-commutative) principal bundle $(C(S^2)\subset C_\theta(S^3))^{C(U(1))}$, it constitutes a proper setup for studying connections (cf.\  \cite{BrzMaj:gau}). In particular we can define a special type of connection, known as a {\em strong connection} \cite{Haj:str}. There are various equivalent ways of defining such connections \cite[Theorem~2.3]{DabGro:str}. One of the possibilities is to say that a strong connection in $(C(S^2)\subset C_\theta(S^3))^{C(U(1))}$, is a left $C(S^2)$-linear, right $C(U(1))$-colinear splitting $s: C_\theta(S^3)\to C(S^2)\otimes C_\theta(S^3)$ of the product $\mu : C(S^2)\otimes C_\theta(S^3) \to C_\theta(S^3)$ such that $s(1) = 1\otimes 1$. Such an $s$ can be defined as follows.  Since $C_\theta(S^3)$ is a $\bf Z$-graded algebra, suffices it to define $s$ on homogeneous elements only. Let
$$
s(f) =f\sum_{m=0}^n\pmatrix{n\cr\\ m} \left\{ \begin{array}{ll}
		(a^{n-m}b^{m})^* \otimes a^{n-m} b^m & \mbox{ if $f$ has degree $n\geq 0$}\\
		a^{n-m}b^{m} \otimes (a^{n-m} b^{m})^* & \mbox{ if $f$ has degree $-n < 0$}
		\end{array}
	\right.
$$
Clearly $s$ is left $C(S^2)$-linear, since $C(S^2)$ consists of all elements of degree $0$. Obviously $s(1) = 1\otimes 1$. The last equation in (\ref{pres}) implies that $\mu \circ s = {\rm id}$, i.e., $s$ is a splitting of the product. Finally, degree counting ensures that $s$ commutes with right coactions, i.e., it is right colinear. Thus we have constructed a strong connection in $(C(S^2)\subset C_\theta(S^3))^{C(U(1))}$. Note that $s$ is determined by a map $\ell : C(U(1)) \to C_\theta(S^3)\otimes C_\theta(S^3)$ given by
$$
\ell (Z^n) = \sum_{m=0}^n \pmatrix{n\cr\\ m}(a^{n-m}b^{m})^* \otimes a^{n-m} b^m, \quad \ell(Z^{*n}) = \sum_{m=0}^n \pmatrix{n\cr\\ m}a^{n-m}b^{m} \otimes (a^{n-m} b^{m})^* .
$$
This map, which should be noted satisfies a number of conditions such as bicovariance, can be taken for another definition of a strong connection. The reader interested in details of the equivalence of these notions of a connection to more familiar notion of a connection form or a gauge field is referred to \cite{DabGro:str} and, in greater generality, to  \cite{BrzHaj:rel}.

Strong connections provide one with a contact point between geometry of quantum principal bundles and non-commutative geometry. In non-commutative geometry, one studies finitely generated projective modules as modules of sections on non-commutative vector bundles. In classical geometry every vector bundle can be viewed as a bundle associated to a principal bundle: the standard fibre is a representation space of the structure (gauge) group. Given a quantum principal bundle or a Hopf-Galois extension $(B\subset P)^H$ ($H$ is a Hopf algebra, $P$ is a right $H$-comodule algebra with the coaction $\Delta_R$ and $B$ are coinvariants of $P$) and a right $H$-comodule (corepresentation) $V$ one identifies sections of the associated bundle with the left $B$-module $\Gamma = {\rm Hom}^H(V,P)$ of right $H$-comodule maps, i.e., maps $\phi:V\to P$ such that $\Delta_R\circ \phi = (\phi\otimes {\rm id})\circ \varrho^V$, where $\varrho^V :V\to V\otimes H$ is the coaction (cf.\ \cite[Theorem~4.3]{Brz:tra}). By \cite[Corollary~2.6]{DabGro:str} if $H$ has a bijective antipode, $(B\subset P)^H$ admits a strong connection, and $V$ is finite dimensional, then $\Gamma$ is a finitely generated projective left $B$-module. Furthermore a strong connection induces a connection in module $\Gamma$, which coincides with the Grassmann or Levi-Civita connection. Such a connection characterises projective module and is uniquely (up to a gauge transformation) determined by a projector, i.e., a matrix $e$ with entries from $B$ such that $e^2 =e$ (cf.\ \cite{CunQui:alg}).

In the case of deformed Hopf fibering $(C(S^2)\subset C_\theta(S^3))^{C(U(1))}$, we can take a  one-dimensional corepresentation $V=\bf C$ of $C(U(1))$, with $\varrho^V(1) = 1\otimes Z$. The resulting $\Gamma$ is a module of sections of a line bundle, and there is an algorithm of how to read the projector out of the strong connection (cf.\  \cite{BrzHaj:rel}). Explicitly, if we write $\ell(Z) = \sum_i l_i\otimes r_i$, then the entries of the projector are $e_{ij} = r_il_j$. Thus the projector corresponding to $s$ above comes out as
$$
e = \pmatrix{aa^* & ab^* \cr \\ ba^* & bb^*} =  \pmatrix{z & x_- \cr \\ x_+  & 1-z}.
$$
This is precisely the projector which describes the well-known Dirac monopole \cite{Lan:dec}. 

\section{Conclusion}
In this short note we have shown that  Matsumoto's non-commutative Hopf fibering fits perfectly into the framework of quantum principal bundles, i.e., it is a Hopf-Galois extension. Furthermore we have constructed a strong connection in this quantum principal bundle, and, by considering the associated line bundle, we have identified this connection with the Dirac magnetic monopole potential. By this means we have obtained a non-commutative geometric description of magnetic monopoles.


\end{document}